\documentstyle[12pt,a4]{article}

\newcommand{\be}{\begin{equation}}
\newcommand{\ee}{\end{equation}}
\newcommand{\bea}{\begin{eqnarray}}
\newcommand{\eea}{\end{eqnarray}}
\newcommand{\bed}{\begin{displaymath}}
\newcommand{\eed}{\end{displaymath}}

\newcommand{\bc}{\begin{center}}
\newcommand{\ec}{\end{center}}

\newcommand{\br}{({\bf r})}
\newcommand{\brp}{({\bf r'})}

\begin{document}

\thispagestyle{empty}

\bc

\vspace{1.5cm}

{\tt To be published in Chemical Physics Letters (1995)}

\vspace{2cm}

{\LARGE  Density functional theory using
an optimized exchange-correlation potential }

\vspace{2cm}

{\Large Tobias Grabo and E.K.U. Gross}

\vspace{1.5cm}

Institut f\"ur Theoretische Physik, Universit\"at W\"urzburg,

Am Hubland, D-97074 W\"urzburg, Germany

\vspace{3cm}

{\bf Abstract}
\ec
We have performed self-consistent calculations for first and second
row atoms using a variant of density-functional theory, the optimized
effective potential method, with an approximation due to Krieger, Li
and Iafrate and a correlation-energy functional developed by Colle and
Salvetti. The mean absolute deviation of first-row atomic ground-state
energies from the exact non-relativistic values is 4.7 mH in our
scheme, as compared to 4.5 mH in a recent configuration-interaction
calculation. The proposed scheme is significantly more accurate than
the conventional Kohn-Sham method while the numerical effort
involved is about the same as for an ordinary Hartree-Fock
calculation.

\vspace{1.5cm}

\pagebreak

%*********************
\section{Introduction}
%*********************

Since the development of modern density-functional theory (DFT) by
Hohenberg, Kohn and Sham \cite{hohe&kohn64, kohn&sham65} the accurate
treatment of exchange and correlation (xc) poses a major
challenge. Although there exist quite a number of approximations for
the xc-energy functional
\cite{gross&dreizler90,parr&yang89,laba&andz91, vosk&wilk&nusa80,
perd&wang92, beck88, lee&yang&parr88},
which give rather good results \cite{john&gill&popl92-1,
john&gill&popl92-2}, the need for further improvement is undisputable.

In conventional DFT the xc energy $E_{xc}$ is approximated by {\em
explicit}\/
functionals of the density $\rho$. The local density approximation (LDA)
\cite{vosk&wilk&nusa80} and the generalized gradient approximations (GGA)
\cite{perd&wang92, beck88, lee&yang&parr88}
fall in this category. However, there exists another approach within
the framework of DFT, the so-called optimized effective potential (OEP)
method first suggested by Sharp and Horton \cite{shar&hort53} and refined by
Talman and Shadwick \cite{talm&shad76}. In this approach, the total
energy (atomic units are used throughout)
\begin{eqnarray} \label{energie}
E_{tot}^{OEP}\left[ \{ \varphi_{j \sigma} \} \right] & = &
 \sum_{\sigma = \uparrow, \downarrow} \sum_{i=1}^{N_{\sigma}} \int
\varphi_{i \sigma}^{\ast}({\bf r})
 \left(-\frac{1}{2} {\bf\nabla}^{2} \right) \varphi_{i \sigma}({\bf r})
 d{\bf r} \nonumber \\
& & - Z \int  \frac{\rho({\bf r})}{\vert {\bf r} \vert } d{\bf r} \nonumber \\
& & +  \frac{1}{2}  \int \int \frac{\rho({\bf r})
       \rho({\bf r'})}{\vert {\bf r - r' \vert}}
       \ d{\bf r} \ d{\bf r'} \nonumber \\
& & + E_{xc}\left[ \{\varphi_{j \sigma} \} \right]
\end{eqnarray}
is a functional of $N = N_{\uparrow} + N_{\downarrow}$ spin orbitals
$\varphi_{j\sigma}$ resulting
from a single-particle Schr\"odinger equation with a local effective
potential:
\be \label{1p-eq}
\left( -\frac{{\bf \nabla}^2}{2} + V_{\sigma}\br \right) \varphi_{j
\sigma} \br = \varepsilon_{j \sigma} \varphi_{j \sigma} \br \qquad j =
1, \ldots, N_{\sigma} \qquad \sigma = \uparrow, \downarrow .
\ee
The {\em optimized}\/ effective potential, $V_{\sigma}^{OEP} \br$, is
determined
by the condition that its orbitals be the ones that minimize the
energy functional (\ref{energie}). The stationarity condition
\be \label{statcond}
\left. \frac{\delta E_{tot}^{OEP}}{\delta V_{\sigma}\br} \right|_{V=V^{OEP}} =
0
\ee
may be written by virtue of the chain rule for functional derivatives as
\begin{equation} \label{derv}
\left. \sum_{i} \int d{\bf r'} \frac{ \delta E_{tot}^{OEP}\left[ \left\{
\varphi_{j \sigma} \right\} \right] }{ \delta
\varphi_{i\sigma}({\bf r'}) }
 \frac{\delta \varphi_{i\sigma}({\bf r'})}{\delta V_{\sigma} ({\bf
r})} \right|_{V=V^{OEP}}
 + c.c.
 = 0.
\end{equation}
Given an approximation of $E_{xc}\left[ \{\varphi_{j \sigma} \}
\right]$, the first derivative in the integrand of equation
(\ref{derv}) is easily computed from equation
(\ref{energie}). To obtain a more accessible form of the second
derivative in equation (\ref{derv}), one may use perturbation theory for
an infinitesimal perturbing potential $\delta V_{\sigma}\br$ to get
\be \label{derv2}
\frac{\delta \varphi_{i\sigma}({\bf r'})}{\delta V_{\sigma} ({\bf r})}
=
- \sum_{\stackrel{k=1}{k \neq i}}^{\infty}
\frac{\varphi_{k\sigma}^{\ast}({\bf r}) \varphi_{k\sigma}({\bf
r'})}{\varepsilon_{k\sigma} - \varepsilon_{i\sigma}} \varphi_{i\sigma}({\bf
r}).
\ee
Insertion of the functional derivative of (\ref{energie}) and
equation (\ref{derv2}) into equation (\ref{derv}), leads, via the
one-particle equation (\ref{1p-eq}), to the following
integral equation:
\begin{equation} \label{OEP-int}
\sum_{i=1}^{N_{\sigma}}  \int  d{\bf  r'} \left( V_{xc
\sigma}^{OEP}({\bf r'}) - u_{xc i \sigma}({\bf r'}) \right)
 \left( \sum_{\stackrel{k=1}{k \neq i}}^{\infty}
 \frac{\varphi_{k\sigma}^{\ast}({\bf r}) \varphi_{k\sigma}({\bf
 r'})}{\varepsilon_{k\sigma} - \varepsilon_{i\sigma} } \right)
 \varphi_{i\sigma}({\bf r}) \varphi_{i\sigma}^{\ast}( {\bf r'} )
 + c.c. = 0
\end{equation}
with
\begin{equation} \label{uxc}
u_{xc i \sigma}({\bf r})
:=
 \frac{ 1 }{ \varphi_{i\sigma}^{\ast} ({\bf r}) }
 \frac{ \delta E_{xc}\left[ \{ \varphi_{j\sigma} \} \right]
}{ \delta \varphi_{i\sigma}({\bf r}) }
\end{equation}
and
\begin{equation}
V_{xc \sigma}^{OEP}({\bf r})
:=
 V_{\sigma}^{OEP}({\bf r}) +
 \frac{Z}{\vert {\bf r} \vert} -
 \int
 \frac{ \rho({\bf r'}) }{ \vert {\bf r} - {\bf r'} \vert }
 d{\bf r'}.
\end{equation}

We emphasize that from a fundamental point of view the total-energy
functional (\ref{energie}) is identical with the Hohenberg-Kohn energy
functional
\be \label{energie2}
E_{tot}^{HK} \left[ \rho_{\uparrow}, \rho_{\downarrow} \right] =
T_{S} \left[ \rho_{\uparrow}, \rho_{\downarrow} \right]
- Z \int  \frac{\rho({\bf r})}{\vert {\bf r} \vert } d{\bf r} \nonumber \\
+  \frac{1}{2}  \int \int \frac{\rho({\bf r})
       \rho({\bf r'})}{\vert {\bf r - r' \vert}}
       \ d{\bf r} \ d{\bf r'} \nonumber \\
+ E_{xc} \left[ \rho_{\uparrow}, \rho_{\downarrow} \right]
\ee
of conventional spin-DFT: By virtue of the Hohenberg-Kohn
theorem for noninteracting systems, the spin orbitals
$\varphi_{j \sigma}$ are functionals of the spin densities
$\rho_{\sigma}$ so that
\be \label{hktoep}
E_{tot}^{HK} \left[ \rho_{\uparrow}, \rho_{\downarrow} \right] =
E_{tot}^{OEP} \left[ \{ \varphi_{j\sigma} \left[ \rho_{\uparrow},
\rho_{\downarrow} \right] \} \right].
\ee
In other words, the orbital functional (\ref{energie}) is an {\em
implicit}\/ density functional. If the (unknown) exact xc functional
were used in equation (\ref{energie}) then the optimized effective
potential determined by equation (\ref{OEP-int}) would be the exact
Kohn-Sham potential. Any approximation of $E_{xc}$ in equation
(\ref{energie}), on the other hand, leads to an approximate Kohn-Sham
potential.

There are three non-trivial density functionals contributing to the
Hohenberg-Kohn-total-energy functional (\ref{energie2}): the
non-interacting kinetic
energy functional $T_{S}\left[ \rho_{\uparrow}, \rho_{\downarrow}
\right]$, the exchange part $E_{x}\left[ \rho_{\uparrow},
\rho_{\downarrow} \right]$  and
the correlation part
$E_{c}\left[ \rho_{\uparrow}, \rho_{\downarrow} \right]$ of
$E_{xc}\left[ \rho_{\uparrow}, \rho_{\downarrow} \right]$. If
$T_{S}\left[ \rho_{\uparrow}, \rho_{\downarrow} \right]$ is
approximated by an LDA, one obtains the Thomas-Fermi model. The
transition from Thomas-Fermi to modern Kohn-Sham theory is equivalent
to replacing the {\em approximate\/} LDA functional by the {\em exact
orbital representation}
\be
T_{S}^{exakt} \left[ \rho_{\uparrow}, \rho_{\downarrow} \right] =
 \sum_{\sigma = \uparrow, \downarrow} \sum_{i=1}^{N_{\sigma}} \int
\varphi_{i \sigma}^{\ast}({\bf r})
 \left(-\frac{1}{2} {\bf\nabla}^{2} \right) \varphi_{i \sigma}({\bf r})
 d{\bf r}.
\ee
The transition from standard Kohn-Sham theory to the OEP method can be
viewed in much the same way: While in ordinary Kohn-Sham theory the
exchange-energy functional is {\em approximated\/} by LDA or GGA-type
functionals such as the one due to Becke \cite{beck88}, the OEP method
allows one to employ the {\em exact orbital representation\/}
\be \label{exhf}
E_{x}^{exact}\left[ \rho_{\uparrow}, \rho_{\downarrow} \right] = -
\frac{1}{2}\sum_{\sigma = \uparrow,
\downarrow} \sum_{j,k}
\int d{\bf r} \int d{\bf r'} \
 \frac{\varphi_{j\sigma}^{\ast}\br \varphi_{k\sigma}^{\ast}\brp
\varphi_{k\sigma}\br
\varphi_{j\sigma}\brp}{\vert {\bf r - r'} \vert}
\ee
of the Kohn-Sham-exchange-energy functional\footnote{The so-called
Kohn-Sham-exchange-energy functional \protect
\cite{enge&vosk93} is distinguished from the Hartree-Fock exchange energy
expression by the fact that the orbitals in (\protect \ref{exhf}) come
from a {\em local\/} single particle potential. It is this restriction
to {\em local\/} potentials that allows one to establish the
Hohenberg-Kohn 1-1 correspondence between densities and potentials
which implies that the orbitals in equation (\protect \ref{exhf}) and
thus $E_{x}$ and $E_{c}$ are implicit functionals of the {\em
densities}. If the
restriction to local potentials is dropped, then only a 1-1 mapping
between (non-local) potentials $w( {\bf r}, {\bf r'})$ and one-particle-density
matrices $\rho( {\bf r}, {\bf r'})$ can be established \protect
\cite{gilb75}. In this formalism {\em Hartree-Fock\/} exchange is
treated {\em exactly\/} but the correlation energy becomes a
functional of the {\em density matrix\/} rather than the density,
which can lead to a $N$-representability problem \protect
\cite{erda&smit87}. For the density, on the other hand,
$N$-representability is always satisfied \protect
\cite{harr78,zumb&masc83}. Therefore the restriction to local potentials and
densities (rather than non-local potentials and density matrices) is
essential.},
 which leads, by virtue of equation (\ref{statcond}), to
the variationally {\em best local\/} exchange potential
$V_{x\sigma}^{OEP} \br$.
Of particular importance is the fact that $V_{x\sigma}^{OEP} \br$ is manifestly
self-inter\-action free. Of course, the correlation part
$E_{c}\left[ \rho_{\uparrow}, \rho_{\downarrow} \right]$ still
has to be approximated, but even for $E_{c}\left[ \rho_{\uparrow},
\rho_{\downarrow} \right]$ the representation in terms of orbitals
allows more flexibility in the construction of approximate
functionals. In particular, $V_{xc\sigma}^{OEP} \br$ is currently the
only approximate xc potential featuring the required discontinuities
as a function of the particle number $N_{\sigma}$
\cite{krie&li&iafr92-1}.

The full OEP
method, however, has a serious drawback: The solution of the integral
equation (\ref{OEP-int}) is numerically very involved.
Recently, Krieger, Li and Iafrate \cite{krie&li&iafr92-1,
krie&li&iafr92-2} have proposed an approximation (which we will refer to as
KLI) to the OEP integral equation (\ref{OEP-int}), in which
$V_{\sigma}^{OEP}\br$ is
essentially obtained from the solution of two linear $(N_{\sigma}
\times N_{\sigma} )$
equations. They also showed that in the x-only case,
the results of the
KLI approximation \cite{krie&li&iafr92-1, krie&li&iafr92-2,
krie&li&iafr93} are nearly
identical with those of the exact x-only OEP-method \cite{norm&koel84,
enge&chev&macd&vosk92, enge&vosk93}. In
addition, the KLI approximation preserves
all of the important advantages of the exact OEP method, such as the
correct asymptotic $-\frac{1}{r}$ decay of $V^{OEP}\br$ and the
discontinuities of $V_{xc\sigma}^{OEP}\br$ as a function of the
particle number.

So far, the KLI approximation has been used exclusively in the x-only
limit. In this letter, we present the first KLI calculations  which
include correlation effects. From the various
available correlation-energy functionals we choose the one developed
by Colle and Salvetti  \cite{coll&sall75, coll&sall79}. This
orbital-dependent
functional gives correlation energies within a few
percent of exact values if Hartree-Fock orbitals are inserted. The
transformed version of it, the density
functional of Lee, Yang and Parr \cite{lee&yang&parr88}, has
been shown to perform very
well in self-consistent calculations \cite{zhu&lee&yang93}, especially
in connection with an exchange energy
functional proposed by
Becke \cite{beck88}.

%***************
\section{Theory}
%***************

In order to find an approximate solution to the OEP integral equation
(\ref{OEP-int}), Krieger, Li and Iafrate started from the Slater
exchange potential \cite{slat51},
given by
\be \label{slatpot}
V_{x \sigma}^{S} \br = \frac{1}{\rho_{\sigma}\br}
\sum_{i=1}^{N_{\sigma}} \rho_{i \sigma}\br u_{xi\sigma} \br,
\ee
with
\be
\rho_{i\sigma} \br := \vert \varphi_{i \sigma} \br \vert^{2}
\ee
and
\begin{equation}
u_{x i \sigma}({\bf r})
:=
 \frac{ 1 }{ \varphi_{i\sigma}^{\ast} ({\bf r}) }
 \frac{ \delta E_{x\sigma}\left[ \{ \varphi_{j\sigma}({\bf r}) \} \right]
}{ \delta \varphi_{i\sigma}({\bf r}) },
\end{equation}
where $E_{x\sigma}$ for $\sigma = \uparrow, \downarrow$ represents the
spin-up and spin-down contribution, respectively, to the right-hand
side of equation (\ref{exhf}).
KLI \cite{krie&li&iafr92-1} argue that a generalized form of this
potential, which differs
from equation (\ref{slatpot})
only through orbital dependent constants $C_{i\sigma}$ added to
$u_{xi\sigma} \br$, should yield a better approximation than Slater's
formula. Moreover, they were able to
show through a series of manipulations \cite{krie&li&iafr92-2}, that
the exact
$V_{xc\sigma}^{OEP} \br$ may be - up to within a small term they neglected -
cast into a form similar to (\ref{slatpot}), namely
\be \label{kli-eq}
 V_{xc\sigma}^{OEP}\br \approx
 V_{xc\sigma}^{KLI}\br =
 \frac{ 1 }{ \rho_{\sigma}({\bf r}) }
 \sum_{i=1}^{N_{\sigma}} \rho_{i\sigma}({\bf r})
 \left[ u_{xci\sigma}({\bf r}) + \left(\bar{V}_{xci\sigma}^{OEP}  -
        \bar{u}_{xci\sigma}  \right) \right]
\ee
where the constants $\left( \bar{V}_{xci\sigma}^{OEP} -
\bar{u}_{xci\sigma} \right)$ are the
solution of the linear equation
\be \label{lin-eq}
\sum_{i=1}^{N_{\sigma}-1} \left( \delta_{ji} - M_{ji\sigma} \right) \left( \bar
V_{xci\sigma}^{OEP} - \bar u_{xci\sigma} \right)
=
\bar V_{xcj\sigma}^{S} - \bar u_{xcj\sigma}
\qquad j= 1, \ldots, N_{\sigma}-1
\ee
with
\be
M_{ji\sigma} := \int d{\bf r} \ \frac{\rho_{j\sigma}\br
\rho_{i\sigma}\br}{\rho_{\sigma}\br},
\ee
\be
V_{xc\sigma}^{S}\br := \sum_{i=1}^{N}
\frac{\rho_{i\sigma}\br}{\rho_{\sigma}\br} u_{xci\sigma}\br.
\ee
$\bar u_{xcj\sigma}$ denotes the average value of
$u_{xcj\sigma}\br$ taken over the density of the $j\sigma$ orbital, i.e.
\be
\bar u_{xcj\sigma} = \int \rho_{j \sigma} \br
u_{xcj\sigma}\br d{\bf r}
\ee
and similarly for $\bar V_{xc\sigma}^{S}$. The indices in equation
(\ref{lin-eq}) run over all occupied orbitals except the one
corresponding to the highest single-particle energy eigenvalue
$\varepsilon_{N\sigma}$. If the highest occupied orbital is degenerate
all orbitals with the energy $\varepsilon_{N\sigma}$  are
excluded from the linear equation (\ref{lin-eq}).

Splitting the exchange-correlation energy $E_{xc}$, as above, into an
exchange and a correlation part, the exchange part (\ref{exhf}) leads
via equation (\ref{uxc}) to the following expression for
$u_{xj\sigma}$:
\be
u_{xj\sigma}\br = - \frac{1}{\varphi_{j\sigma}^{\ast}\br} \sum_{k}
\int d{\bf r'}
\ \frac{\varphi_{j\sigma}^{\ast}\brp \varphi_{k\sigma}\brp
\varphi_{k\sigma}^{\ast}\br}{\vert {\bf r - r'} \vert}.
\ee

The correlation part, $E_{c}$, within the approximation by Colle and
Salvetti \cite{coll&sall75, coll&sall79}, is given by \cite{lee&yang&parr88}
\bea \label{csec2}
E_{c} & = & \  - \ ab \int \gamma \br \xi \br \Biggl[ \sum_{\sigma}
\rho_{\sigma} \br \sum_{i} \mid \!{\bf \nabla} \varphi_{i\sigma} \br
\! \mid^{2}
\ - \ \ \frac{ab}{4}  \mid \! {\bf \nabla} \rho \br \! \mid^{2} \nonumber \\
&&  \qquad \qquad \qquad \qquad \ - \ \frac{ab}{4}  \sum_{\sigma}
\rho_{\sigma}\br \triangle \rho_{\sigma} \br \
+ \ \frac{ab}{4}  \rho \br \triangle \rho \br \Biggr] d{\bf r} \nonumber \\
&& -a \int \gamma \br \frac{ \rho \br }{\eta \br} \ d{\bf r},
\eea
where
\begin{eqnarray}
\label{gam}
\gamma \br& = & 4 \ \frac{\rho_{\uparrow}\br
\rho_{\downarrow}\br}{\rho \br^{2}}, \\
\label{eta}
\eta \br & = & 1 + d \rho \br^{-\frac{1}{3}},\\
\label{xsi}
\xi \br & = & \frac{\rho \br^{-\frac{5}{3}} e^{-c \rho
\br^{-\frac{1}{3}}}}{\eta \br}.
\end{eqnarray}
The constants $a$, $b$, $c$ and $d$ are given by
\bed
\begin{array}{ll}
a = 0.04918, \qquad & b = 0.132, \\
c = 0.2533, & d = 0.349 .
\end{array}
\eed
Performing the functional derivative with respect to the one-particle
orbitals, one obtains for $u_{cj\sigma}\br$
%\pagebreak
\bea
u_{cj \sigma} \br & =  & - \frac{a}{\eta \br} \left( \gamma \br + \rho \br
\frac{1}{\varphi_{j \sigma}^{\ast} \br} \frac{\partial \gamma
\br}{\partial \varphi_{j \sigma} \br} \right) - \frac{ad}{3} \gamma
\br \frac{\rho \br^{-\frac{1}{3}}}{\eta \br^{2}} \nonumber \\
& & - \frac{ab}{4} \frac{1}{\varphi_{j \sigma}^{\ast} \br} \left[
\frac{\partial}{\partial \varphi_{j \sigma} \br} \Bigl( \gamma \br \xi
\br \Bigr) \right] \Biggl[  4 \sum_{\sigma'}
\rho_{\sigma'} \br \sum_{i} \mid \! {\bf \nabla} \varphi_{i \sigma'} \br
\! \mid^{2}
\nonumber \\
& & \qquad \qquad \qquad \quad \
- \Bigl( {\bf \nabla} \rho \br \Bigr)^{2}
+ \Bigl(\rho_{\uparrow} \br \triangle \rho_{\downarrow}\br +
\rho_{\downarrow} \br  \triangle \rho_{\uparrow}\br \Bigr) \Biggr]
 \nonumber \\
& & - \frac{ab}{2}\ {\bf \nabla} \Bigl( \gamma \br \xi \br \Bigr) \Bigl(
{\bf \nabla} \rho \br + {\bf \nabla} \rho_{\tilde \sigma} \br \Bigr) \nonumber
\\
& & - \frac{ab}{4}\, \triangle \Bigl( \gamma \br \xi \br \Bigr)
\rho_{\tilde \sigma} \br \nonumber \\
& & - \ ab \ \gamma \br \xi \br \left[ \sum_{i} \mid \! {\bf \nabla}
\varphi_{j \sigma} \br \! \mid^{2} + \frac{1}{2} \Bigl( \triangle \rho
\br + \triangle \rho_{\tilde \sigma} \br \Bigr) \right] \nonumber \\
& & + \ ab \ \frac{{\bf \nabla} \varphi_{j \sigma}^{\ast}\br }{\varphi_{j
\sigma}^{\ast} \br}   {\bf \nabla} \Bigl( \gamma \br
\xi \br  \rho_{\sigma} \br \Bigr) \nonumber \\
& & + \  ab  \ \frac{\triangle \varphi_{j \sigma}^{\ast} \br }{\varphi_{j
\sigma}^{\ast} \br} \rho_{\sigma} \br \gamma \br \xi \br
\eea
where $\tilde \sigma$ denotes the spin projection opposite to
$\sigma$, i.e. $\tilde \sigma = \uparrow$ if $\sigma = \downarrow$ and vice
versa.

In the numerical calculations, equation (\ref{kli-eq}) is solved
self-consistently together with the local one-particle equation
\be \label{1T-eq}
\left( -\frac{1}{2} {\bf \nabla}^{2}  +  V_{\sigma}^{OEP}({\bf  r})
\right)
\varphi_{j\sigma}({\bf r}) = \varepsilon_{j\sigma}
\varphi_{j\sigma}({\bf r})
\ee
where
\begin{equation}
V_{\sigma}^{OEP}({\bf r})
:=
 V_{xc\sigma}^{KLI}({\bf r}) -
 \frac{Z}{\bf r} +
 \int
 \frac{ \rho({\bf r'}) }{ \vert {\bf r} - {\bf r'}
 \vert }
 d{\bf r'}.
\end{equation}

%****************
\section{Results}
%****************

Equation (\ref{1T-eq}) is solved by expanding the single-particle
orbitals in spherical harmonics. The resulting equation for the radial
part of the wave functions is then solved numerically.
The main features of the code are as described in Ref. 28.

For comparison, we have also performed calculations using the conventional
Kohn-Sham method with two of the best standard exchange-correlation energy
functionals. The first one of these is the exchange-energy functional
by Becke \cite{beck88} combined with the correlation-energy functional
by Lee, Yang and Parr \cite{lee&yang&parr88}, referred to as BLYP. The
other is the generalized gradient approximation by Perdew and Wang
\cite{perd&wang92}, referred to as PW91.

\begin{table}[phbt]
\caption{ \sl Total absolute ground-state energies for first-row atoms
from various self-consistent calculations. CI values from
Ref. 29. $\bar \triangle$ denotes the mean absolute deviation from the
exact nonrelativistic values [30]. All numbers in Hartree units. }
\label{tabetot1}
\begin{center}
\begin{tabular}{|l|r|r|r|r|r|}
\hline
 & present & BLYP & PW91 & $\mbox{CI   }$ & EXACT \\
\hline
He & 2.9033 & 2.9071 & 2.9000 & 2.9049 & 2.9037 \\
Li & 7.4829 & 7.4827 & 7.4742 & 7.4743 & 7.4781 \\
Be & 14.6651 & 14.6615 & 14.6479 & 14.6657 & 14.6674 \\
B & 24.6564 & 24.6458 & 24.6299 & 24.6515 & 24.6539 \\
C & 37.8490 & 37.8430 & 37.8265 & 37.8421 & 37.8450 \\
N & 54.5905 & 54.5932 & 54.5787 & 54.5854 & 54.5893 \\
O & 75.0717 & 75.0786 & 75.0543 & 75.0613 & 75.067 \\
F & 99.7302 & 99.7581 & 99.7316 & 99.7268 & 99.734 \\
Ne & 128.9202 & 128.9730 & 128.9466 & 128.9277 & 128.939 \\
$\bar \triangle$ & 0.0047 & 0.0108 & 0.0114 & 0.0045 & \\
\hline
\end{tabular}
\end{center}
\end{table}

\begin{table}[phbt]
\caption{\sl Total absolute ground-state energies for second-row atoms
from various self-consistent calculations.  $\bar \triangle$ denotes the
mean absolute deviation from Lamb-shift corrected
experimental values, taken from Ref. 3. All numbers in Hartree units.}
\label{tabetot2}
\begin{center}
\begin{tabular}{|l|r|r|r|r|}
\hline
 & present & BLYP & PW91 & EXPT \\
\hline
Na & 162.256 & 162.293 & 162.265 & 162.257 \\
Mg & 200.062 & 200.093 & 200.060 & 200.059 \\
Al & 242.362 & 242.380 & 242.350 & 242.356 \\
Si & 289.375 & 289.388 & 289.363 & 289.374 \\
P & 341.272 & 341.278 & 341.261 & 341.272 \\
S & 398.128 & 398.128 & 398.107 & 398.139 \\
Cl & 460.164 & 460.165 & 460.147 & 460.196 \\
Ar & 527.553 & 527.551 & 527.539 & 527.604 \\
$\bar \triangle$ & 0.013 & 0.026 & 0.023 & \\
\hline
\end{tabular}
\end{center}
\end{table}

Table \ref{tabetot1} shows the total absolute
ground-state energies of the first-row atoms. For these atoms, there
exist accurate estimates of the exact non-relativistic values obtained
from experimental ionisation energies and improved {\em ab initio}\/
calculations by Davidson
et al. \cite{davi&hags&chak&uma91}. It is evident from the Table,
that the density functional methods perform quite
well. The mean absolute errors, $\bar
\triangle$, given in the last row of Table \ref{tabetot1}, clearly
show that the present approach is significantly more accurate than the
conventional Kohn-Sham method and nearly as accurate as recent CI
results by Montgomery et al. \cite{mont&ocht&pete94}.
The situation is similar for second-row atoms, as can be seen from
Table \ref{tabetot2}. As the
relativistic effects for these atoms are more important and experiments
increasingly difficult, the comparison of the calculated values with
the Lamb-shift corrected experimental ones (from Ref. 3)
has to be done
cautiously and is by no means as rigorous as for first-row
atoms. Nevertheless, our calculated values seem to mirror these
experimental values more closely than the other approximations.

\begin{table}
\caption{ \sl Exchange and correlation energies from various
approximations. All values in Hartree units.} \label{tabxc}
\begin{center}
\begin{tabular}{|l|r|r|r||r|r|r|}
\hline
 & $-E_{x}^{present}$ & $-E_{x}^{BLYP}$ & $-E_{x}^{PW91}$ &
$-E_{c}^{present}$ & $-E_{c}^{BLYP}$ & $-E_{c}^{PW91}$\\
\hline
He & 1.028 & 1.018 & 1.009 & 0.0416 & 0.0437 & 0.0450 \\
Li & 1.784 & 1.771 & 1.758 & 0.0509 & 0.0541 & 0.0571 \\
Be & 2.674 & 2.658 & 2.644 & 0.0934 & 0.0954 & 0.0942 \\
B & 3.760 & 3.727 & 3.711 & 0.1289 & 0.1287 & 0.1270 \\
C & 5.064 & 5.028 & 5.010 & 0.1608 & 0.1614 & 0.1614 \\
N & 6.610 & 6.578 & 6.558 & 0.1879 & 0.1925 & 0.1968 \\
O & 8.200 & 8.154 & 8.136 & 0.2605 & 0.2640 & 0.2587 \\
F & 10.025 & 9.989 & 9.972 & 0.3218 & 0.3256 & 0.3193 \\
Ne & 12.110 & 12.099 & 12.082 & 0.3757 & 0.3831 & 0.3784 \\
Na & 14.017 & 14.006 & 13.985 & 0.4005 & 0.4097 & 0.4040 \\
Mg & 15.997 & 15.986 & 15.967 & 0.4523 & 0.4611 & 0.4486 \\
Al & 18.081 & 18.053 & 18.033 & 0.4905 & 0.4979 & 0.4891 \\
Si & 20.295 & 20.260 & 20.238 & 0.5265 & 0.5334 & 0.5322 \\
P & 22.649 & 22.609 & 22.587 & 0.5594 & 0.5676 & 0.5762 \\
S & 25.021 & 24.967 & 24.944 & 0.6287 & 0.6358 & 0.6413 \\
Cl & 27.530 & 27.476 & 27.453 & 0.6890 & 0.6955 & 0.7055 \\
Ar & 30.192 & 30.139 & 30.116 & 0.7435 & 0.7515 & 0.7687 \\
\hline
\end{tabular}
\end{center}
\end{table}

\begin{table}
\caption{\sl Total absolute exchange energies of spherical first and second row
atoms for various self consistent {\em x-only}\/ calculations. The
exact OEP data are from Refs. 22 and 23. All values in
Hartree units.} \label{tabxonly}
\begin{center}
\begin{tabular}{|l|r|r|r|r|r|}
\hline
 & KLI & B88 & PW91 & OEP \\
\hline
He  & 1.026 & 1.016 & 1.005 & 1.026 \\
Li & 1.781 & 1.768 & 1.754 & 1.781 \\
Be & 2.667 & 2.652 & 2.638 & 2.666 \\
N & 6.603 & 6.569 & 6.547 & 6.604  \\
Ne & 12.099 & 12.086 & 12.061 & 12.105 \\
Na & 14.006 & 13.993 & 13.968 & 14.013  \\
Mg & 15.983 & 15.972 & 15.950 & 15.988 \\
P & 22.633 & 22.593 & 22.565 & 22.634 \\
Ar & 30.174 & 30.122 & 30.089 & 30.175  \\
\hline
\end{tabular}
\end{center}
\end{table}

For further analysis, we list, in Table \ref{tabxc}, the values of
$E_{x}$ and $E_{c}$ separately. The data show two main features:
First, the results
for $E_{x}$ are lowest for our method and highest
for PW91, while the BLYP-values lie somewhere in between. And second,
for $E_{c}$, this trend is reversed, as now our results are
highest and the ones from BLYP and PW91 are lower in nearly all
cases. In Table \ref{tabxonly} we show results of various {\em
x-only}\/ calculations performed with only the exchange-energy parts of the
respective functionals. For the spherical atoms listed, there exist
exact x-only OEP values
\cite{enge&chev&macd&vosk92,enge&vosk93}. It is evident,
that the KLI-approximation gives values much closer to the exact
ones than the generalized gradient approximations.
{}From this and from Table \ref{tabxc} one may conclude that in the BLYP
and PW91 schemes an error cancellation between exchange and
correlation energies occurs which leads to rather good total
energies. Exchange and correlation energies {\em separately}, however,
are poorly reproduced in the BLYP and PW91 approaches. By contrast, in
our scheme, both exchange and correlation energies are of high
quality.

\begin{table}[htb]
\caption{ \sl Ionisation potentials calculated from ground-state-energy
differences of neutral atoms. CI values are from Ref. 29. $\bar
\triangle$ denotes the
mean absolute deviation from the experimental values, taken from Ref.
31. All values in Hartree units.} \label{tabipot}
\begin{center}
\begin{tabular}{|c|r|r|r|r|r|}
\hline
 & present& BLYP & PW91 & $\mbox{CI   }$ & EXPT \\
\hline
He & 0.903 & 0.912 &  & 0.905 & 0.903 \\
Li & 0.203 & 0.203 & 0.207 & 0.198 & 0.198 \\
Be & 0.330 & 0.330 & 0.333 & 0.344 & 0.343 \\
B & 0.314 & 0.309 & 0.314 & 0.304 & 0.305 \\
C & 0.414 & 0.425 & 0.432 & 0.413 & 0.414 \\
N & 0.527 & 0.542 & 0.551 & 0.534 & 0.534 \\
O & 0.495 & 0.508 & 0.505 & 0.499 & 0.500 \\
F & 0.621 & 0.656 & 0.660 & 0.639 & 0.640 \\
Ne & 0.767 & 0.808 & 0.812& 0.792 & 0.792 \\
$\bar \triangle$ & 0.009 & 0.010 & 0.014 & 0.001 & \\
\hline
Na & 0.191 & 0.197 & 0.198 & & 0.189 \\
Mg & 0.275 & 0.280 & 0.281 & & 0.281 \\
Al & 0.218 & 0.212 & 0.221 & & 0.220 \\
Si & 0.294 & 0.294 & 0.305 & & 0.300 \\
P & 0.379 & 0.376 & 0.389 & & 0.385 \\
S & 0.380 & 0.379 & 0.379 & & 0.381 \\
Cl & 0.471 & 0.476 & 0.482 & & 0.477 \\
Ar & 0.575 & 0.576 & 0.583 & & 0.579 \\
$\bar \triangle$ & 0.004 & 0.005 & 0.004 & & \\
\hline
\end{tabular}
\end{center}
\end{table}

\begin{table}
\caption{ \sl Ionisation potentials from the highest occupied
 orbital energy of neutral atoms. Experimental values from
Ref. 31. All values in
Hartree units.} \label{tabipot2}
\begin{center}
\begin{tabular}{|c|r|r|r|r|r|}
\hline
 & present & BLYP & PW91 & EXPT \\
\hline
He & 0.945 & 0.585 & 0.583 & 0.903 \\
Li & 0.200 & 0.111 & 0.119 & 0.198 \\
Be & 0.329 & 0.201 & 0.207& 0.343 \\
B & 0.328 & 0.143 & 0.149 & 0.305 \\
C & 0.448 & 0.218 & 0.226 & 0.414 \\
N & 0.579 & 0.297 & 0.308 & 0.534 \\
O & 0.559 & 0.266 & 0.267 & 0.500 \\
F & 0.714 & 0.376 & 0.379 & 0.640 \\
Ne & 0.884 & 0.491 & 0.494 & 0.792 \\
Na & 0.189 & 0.106 & 0.113 & 0.189 \\
Mg & 0.273 & 0.168 & 0.174 & 0.281 \\
Al & 0.222 & 0.102 & 0.112 & 0.220 \\
Si & 0.306 & 0.160 & 0.171 & 0.300 \\
P & 0.399 & 0.219 & 0.233 & 0.385 \\
S & 0.404 & 0.219 & 0.222 & 0.381 \\
Cl & 0.506 & 0.295 & 0.301 & 0.477 \\
Ar & 0.619 & 0.373 & 0.380 & 0.579 \\
\hline
\end{tabular}
\end{center}
\end{table}

\begin{table}
\caption{ \sl Self-consistent electron affinities, (a) from
ground-state-energy differences, and (b) from the highest occupied
orbital energies of the negative ions. CI values are from Ref. 29,
experimental ones from Ref. 31. All values in
Hartree units.} \label{tabea}
\begin{center}
\begin{tabular}{|c|r|r|r|r|}
\hline
 & $\mbox{present}^{a}$ & $\mbox{present}^{b}$ & $\mbox{CI   }$& EXPT\\
\hline
Li & 0.016 & 0.024 & 0.023 & 0.023 \\
B  & -0.002 & 0.033 & 0.008 & 0.010 \\
C & 0.028 & 0.083 & 0.045 & 0.046 \\
O & 0.017 & 0.110 & 0.052 & 0.054 \\
F & 0.082 & 0.208 & 0.125 & 0.125\\
Na & 0.015 & 0.022 & & 0.020 \\
Al & 0.007 & 0.024 & & 0.016 \\
Si & 0.040 & 0.065 & & 0.051 \\
P & 0.022 & 0.048  & & 0.027 \\
S & 0.065 & 0.106 & & 0.076 \\
Cl & 0.122 & 0.174 & & 0.133 \\
\hline
\end{tabular}
\end{center}
\end{table}

Limitations of the three DFT approaches studied here become evident for
ionisation potentials and electron affinities. In Table
\ref{tabipot} we show ionisation potentials calculated from
ground-state-energy differences and CI values from Ref. 29
as well as experimental ones from Ref. 31.
The performance of the three DFT methods is about
the same, while the CI approach leads to clearly better
results. Somewhat surprisingly, the DFT methods work better for the
second-row than for the first-row atoms.

In {\em exact\/} DFT, the highest occupied orbital energy of the neutral
atom is identical with the ionisation potential, while for negative
ions the highest occupied energy level coincides with the electron
affinity of the neutral atom \cite{almb&bart85}. How well ionisation
potentials and
electron affinities are reproduced by the highest occupied energy
eigenvalues resulting
from an {\em approximate\/} xc functional is therefore a measure of the
quality of the functional.
Table \ref{tabipot2} shows the ionisation potentials obtained from the
highest occupied single-particle-energy eigenvalue of the neutral atoms.
The resulting values are worse than
the ones in Table \ref{tabipot}. While the deviation from
experiment resulting from the
BLYP and PW91 functionals is around 100 percent for all atoms, it is
at most 10 percent in our approach. This is due to the fact that in
our approach the xc potential has the correct
$-\frac{1}{r}$ asymptotic behaviour for large $r$ which is not
properly reproduced with the BLYP and PW91 functionals.

For
electron affinities, the situation is much worse, as may be seen from
Table \ref{tabea}. First of all, because of the wrong asymptotic behaviour
of the xc potential for large $r$, there is no convergence for
negative ions within
the self-consistent BLYP and PW91 schemes.
This is not the case for our approach. However, the
resulting electron affinities obtained either from ground-state-energy
differences or from the highest orbital energies of negative ions
are not very reliable, in fact, for the Boron atom, the wrong sign is
obtained from the ground-state-energy differences. Here,
 quantum-chemical approaches, such as CI,
are clearly superior.

%********************
\section{Conclusions}
%********************

We have studied a novel density-functional scheme combining the KLI
approximation for the optimized effective potential with the
correlation-energy functional of Colle and Salvetti. The total
ground-state energies obtained with this method for first and
second-row atoms are significantly more accurate than the ones from
the standard
BLYP and PW91 schemes and only slightly less accurate than recent CI
calculations. Error cancellations between exchange and correlation
energies, well known to be present in the standard LDA, BLYP and PW91
functionals are not found for the new scheme. Therefore,  both
exchange and correlation energies are separately of high accuracy.

The calculated atomic ionisation potentials are in satisfactory
agreement with experiment but clearly inferior to CI results. If the
ionisation potentials are calculated by taking ground-state-energy
differences, the performance of the three DFT schemes is about the
same. If, on the other hand, the ionisation potentials are calculated
from the highest occupied orbital energy the new scheme is clearly
superior to the BLYP and PW91 schemes. Percentage deviations are
better by roughly an order of magnitude. This can be explained by the
fact that the KLI approximation preserves the correct asymptotic
$-\frac{1}{r}$ behaviour of the exact Kohn-Sham exchange potential
while the BLYP and PW91 functionals are deficient in this respect. For
the same reason, negative ions are not bound in the BLYP and PW91
self-consistent schemes so that electron affinities cannot be
obtained, whereas the new scheme allows the calculation of these
quantities. The calculated values, however, are not very reliable in
comparison with experimental data and CI results. This clearly shows
the need of further improving the correlation-energy functional. Work
along these lines is in progress.

We finally emphasize that the numerical effort involved in the
proposed scheme is only marginally higher than that of an ordinary
Hartree-Fock calculation, whereas all the principles of standard DFT
are preserved.
We therefore expect that our approach can be
used with great success in chemical studies of more complex and larger
systems, improving the results
\cite{john&gill&popl92-1,john&gill&popl92-2}  obtained with the
ordinary Kohn-Sham method.

%*************************
\section{Acknowledgements}
%*************************

We gratefully appreciate the help of Dr. E. Engel especially for
providing us with a conventional
Kohn-Sham computer code and
for some helpful discussions. We
would also like to thank Professor J. Perdew for providing us with the PW91
xc subroutine, and Professor
H. Stoll and M. Petersilka
for many useful discussions. This work
was supported in part by the Deutsche Forschungsgemeinschaft.

%**************************

\end{document}